The Naked-eye Optical Transient OT 120926

Yue Zhao
Department of Physics, Lanzhou University, Lanzhou, China, and
Department of Physics and Astronomy, York University, Toronto, Ontario, Canada

Patrick B. Hall, Paul Delaney, J. Sandal
Department of Physics and Astronomy, York University, Toronto, Ontario, Canada

Abstract:

A previously unknown optical transient has been observed in the constellation Bootes. The transient flared to brighter than 5th magnitude, which is comparable to the visual magnitudes of the nearby stars π Bootes and o Bootes. This article describes the relative astrometry and photometry work we have done regarding the transient.

Introduction:

Distant astronomical sources normally invisible to the naked eye which transiently brighten by more than a magnitude to naked-eye visibility (V<6) are of considerable scientific interest. Specific examples include a flare on the Be star HD 160202 (peak V~1, Bakos 1968), SN 1987A (peak V=2.96, Hamuy et al. 1988), GRB 080319B (peak V=5.3, Cwiok et al. 2008, Bloom et al. 2009), and possibly OT 060420 (apparent peak V=4.7, Shamir & Nemiroff 2006). Cataclysmic variable eruptions and flares on M dwarf stars can also in principle create naked-eye transients. M dwarf flares can have peak brightenings, in units of magnitudes of $\Delta V=6$ (Stelzer et al. 2006, Kowalski et al. 2013) and even $\Delta V=9$ (Stanek et al. 2013, Schmidt et al. 2013) or $\Delta B=9.5$ (Schaefer 1990). Cataclysmic variables such as classical novae or dwarf novae of the WZ Sge subtype can brighten by up to $\Delta V=7.5$ magnitudes (Harrison et al. 2004). Superflares on main sequence stars usually have $\Delta V<1$ magnitude (Schaefer et al. 2000), though examples of increases up to $\Delta V=7$ are known (Schaefer 1989). Superflares may be due to reconnection events between the magnetic fields of the star and a close-in giant planet (Rubenstein & Schaefer 2000, Rubenstein 2001).

In part because of the rarity of reports of naked-eye transients, current limits on the rate of their occurrence on the sky are not strong (Shamir & Nemiroff 2009). Current and future large-area sky surveys will reveal more and more of these bright transients. Until then, studies of the transient naked-eye sky must rely on whatever images are available.

At York University, students in the introductory Natural Sciences course on Astronomy have taken photographs of assigned constellations each fall for over a decade. Here we report the details of our discovery (Zhao et al. 2013) of a naked-eye transient in photographs taken by one of those students.

Discovery:

The constellation Bootes was observed from Brampton, Ontario, Canada (79.7667 W, 43.6833 N) by J. Sandal on the evening of 2012 September 25 local time (MJD 56195; UTC 2012 September 26) using a Sony DSC-W570 18.2 Mpix handheld digital camera. These unfiltered observations reveal an optical transient (hereafter referred to as OT 120926) flaring to brighter than 5th magnitude, close in brightness to $\pi$ Bootes and $o$ Bootes. The transient's altitude was between 20 and 30 degrees at the times of observation.

There are two photographs taken on the night of observation suitable for analyzing the properties of the transient. Image DSC1875 (Figure 1) had an exposure time of 0.125 seconds, and stars in it possess a slightly elongated point spread function with a FWHM of about 10 pixels. Image DSC1861 had an exposure time of 2 seconds, and stars in it appear as complicated trails (spanning approximately 13 by 18 pixels) due to camera motion (Figure 2). Information on the two images is given in Table 1; the pixel scale was determined from the separation of $\pi$ Boo and o Boo. OT 120926, $\pi$ Boo and o Boo are approximately equally spaced in a straight line on the sky with $\pi$ Boo in the middle and OT 120926 to the west, as seen in the close-up images in Figure 3.

A third photograph (DSC1864), taken in between the other two at 00:23:14 (UTC), does also show the transient to be present. However, its poor point spread function and low resolution of about 300 arc seconds per pixel makes it useless for detailed analysis. All we can conclude from that image is that the transient is not brighter than $\pi$ Boo in it. All three images of this transient are available on nova.astrometry.net by searching for "Bootes flare".

The two pictures taken at different time with different rotation angles of the camera on the sky show that OT 120926 has an unchanged position relative to $\pi$ Boo and o Boo. That excludes the possibility that OT 120926 is a reflection of Arcturus: if OT 120926 was such an artifact, it would have appeared at a different location on the sky when the camera was rotated. Including the transient, eighteen objects are visible in DSC1861 (one very faintly) and fourteen in DSC1875. Inspection of the images reveals no other objects which cannot be identified with known stars.

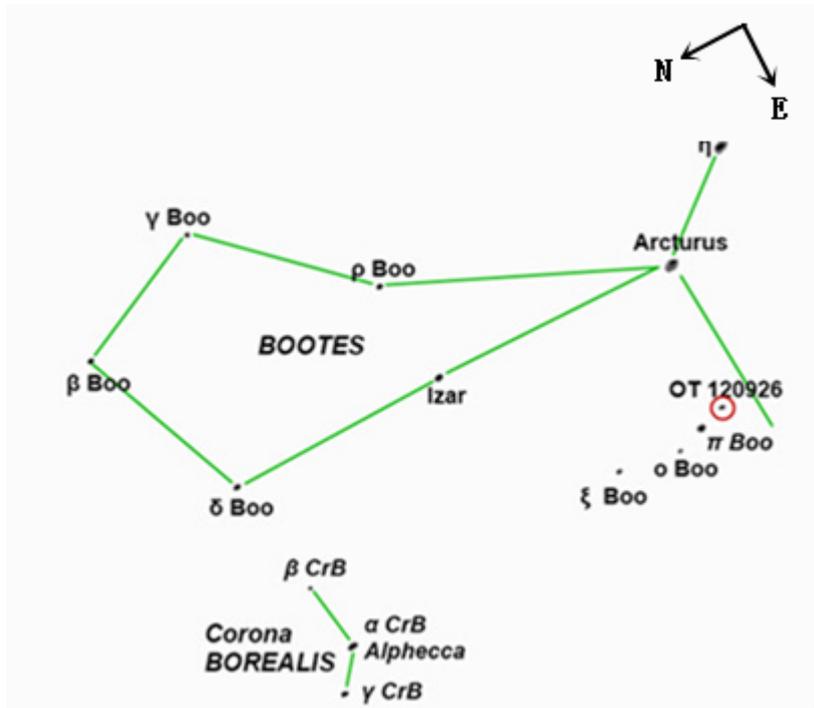

Figure 1: Unfiltered image DSC1875, with constellations sketched. Stars visible in the photo are annotated. The labeled circle indicates the position of OT 120926 near the right edge of the figure.

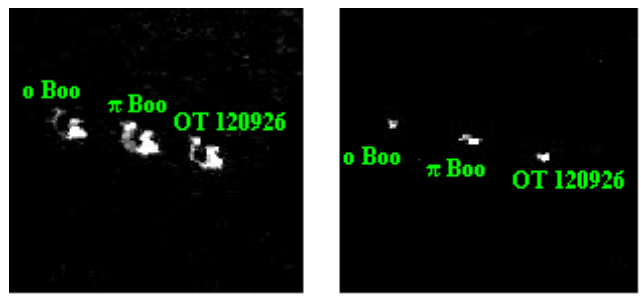

Figure 2: Both unfiltered images above show o Boo (left), π Boo (middle) and OT 120926 (right). The left image is the original version of DSC1861, in which each object has a complicated trail by virtue of camera motion. The region occupied by the trail is approximately 13 x 18 pixels in size. The right image is a processed version in which the unresolved peak at one end of the trail has been isolated.

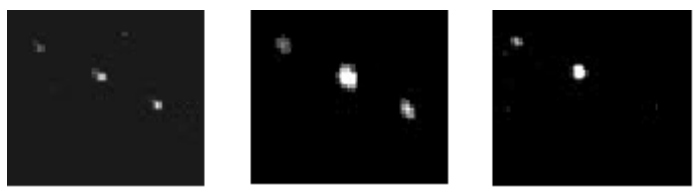

Figure 3: We show o Boo, π Boo and OT 120926 in our two unfiltered images side by side (DSC1861 on the left and DSC1875 in the middle), rotated and rebinned to the same pixel scale, with north up and east left. The third image of the same region (right) is from the Palomar Observatory Sky Survey blue plate, magnified to the same scale.

All Sky Camera Searches:

We searched for the transient on images from the all sky camera operated by The Liverpool Telescope Project (Steele et al. 2004) on the island of La Palma, Spain. We scrutinized images taken between 19:41:14 and 20:05:14 on September 25th 2012 (UTC), which is several hours before the transient was observed in Ontario. Because of moonlight, we smoothed, shifted and combined several images together before seeking around Bootes for the transient. We did not find any object at the anticipated position with brightness comparable to o Boo and π Boo, from which we concluded that those images were taken before the transient brightened to the magnitude observed later.

We also examined images from other all sky cameras. Images from the West Acton Observatory in Acton, MA, spanning the time of the transient's appearance in our images do not show the transient. However, its brightness is limited to V>2.2 (fainter than α CrB) at best. Images are also available from the Kitt Peak National Observatory and MMT Observatory (Pickering 2006) beginning approximately UTC 02:00 on September 26th 2012. They do not show the transient, but the limit on its brightness is only V>2.2 from Kitt Peak and worse on the lower-resolution MMTO images.

Scrutiny of the Liverpool all sky camera images taken the next night (UTC September 26th 2012 between 19:40:13 and 20:04:15) also shows no sign of the transient.

Position Measurement:

In image DSC1861, the stars have the shape of a seeing disk trailed in a complicated pattern due to camera motion. We used IRAF (Image Reduction and Analysis Facility) task "imedit" to isolate only the sharpest part of each object trail, which is shown in Figure 2.

We calculated the coordinates of OT 120926 via a local linear transformation between pixel and equatorial coordinate systems, using the known coordinates of π Boo and o Boo. Our approach assumes the two coordinate systems are locally Cartesian near π Boo, with the equatorial system having only a rotation angle and a scale difference relative to the pixel system. We solved for the coordinates of OT 120926 on the processed image DSC1861 and the image DSC1875 separately using the IRAF task "geomap" with the parameter "fitgeo=rscale". These coordinates and their weighted average are given in Table 2.

Magnitude of the Transient:

On our unfiltered images, the transient OT 120926 was of similar brightness to o Boo but not as bright as π Boo (which is a blended binary). o Boo and π Boo have nearly identical V-band magnitudes (V=4.61 and V=4.51, respectively), but in the B band o Boo (B=5.56) is fainter than π Boo (B=4.59).

The unfiltered relative magnitudes of o Boo and π Boo in our images are consistent with the camera responding to the average flux in the B and V bands. We used those average fluxes with the V band zeropoint to calculate "BV" magnitudes for o Boo and π Boo (4.98 and 4.55, respectively). In both of our images, we measured the magnitude of OT 120926 relative to o Boo and π Boo and averaged to obtain the "BV" magnitude and associated RMS uncertainty of OT 120926 shown in Table 3.

The decline of about one magnitude in the half an hour between our images is less steep than observed in gamma-ray bursts (Bloom et al. 2009). The decline is consistent with the range seen in flares on M dwarfs during the gradual decay phase after peak brightness (Kowalski et al. 2013). If OT 120926 was an M dwarf flare, a fast rise phase may have been missed wherein it could have up to 2.5 magnitudes brighter than in our first image.

We obtain an upper limit to the magnitude of the transient approximately four hours before its detection based on an average image from the Liverpool all sky camera. We shifted 10 images to align π Boo in all of them, subtracted a heavily smoothed version of each image from itself, and then averaged those background-subtracted images together. OT 120926 is not detected in the average image, and because o Boo is the faintest object detected on the average image, the magnitude of OT 120926 must be larger than that of o Boo. This magnitude limit is shown in Figure 4 with an error bar extending off the plot, while the two measured magnitudes of OT 120926 are shown as points with error bars.

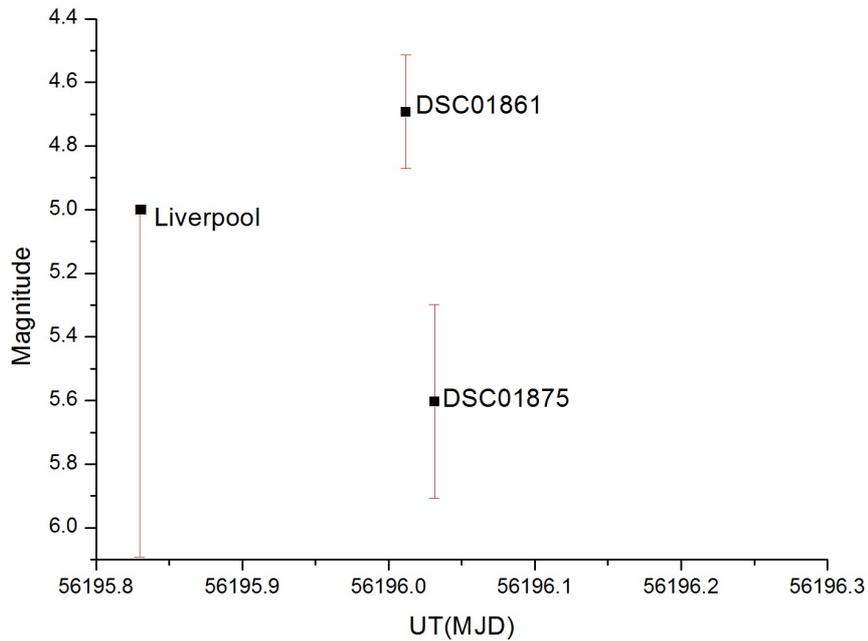

Figure 4: The magnitude of the transient at three different times. Note that the downward error bar on the Liverpool point means the transient's magnitude can be any value larger than 5.

Candidate Identifications for OT 120926 in Quiescence:

No known variable star is listed in the General Catalog of Variable Stars (Samus et al. 2010) within one degree of the transient. There are 3 stars that may be responsible for the transient when we search for candidates on SIMBAD centered on the calculated coordinates within a radius of 5 arcmin. Information on these 3 candidates, plus one other identified in the SDSS, are listed in Table 4, including their distance from OT 120926 in units of the random uncertainty (sigma) on the position of OT 120926. Objects labeled in Table 3 are shown in a finding chart in Figure 5.

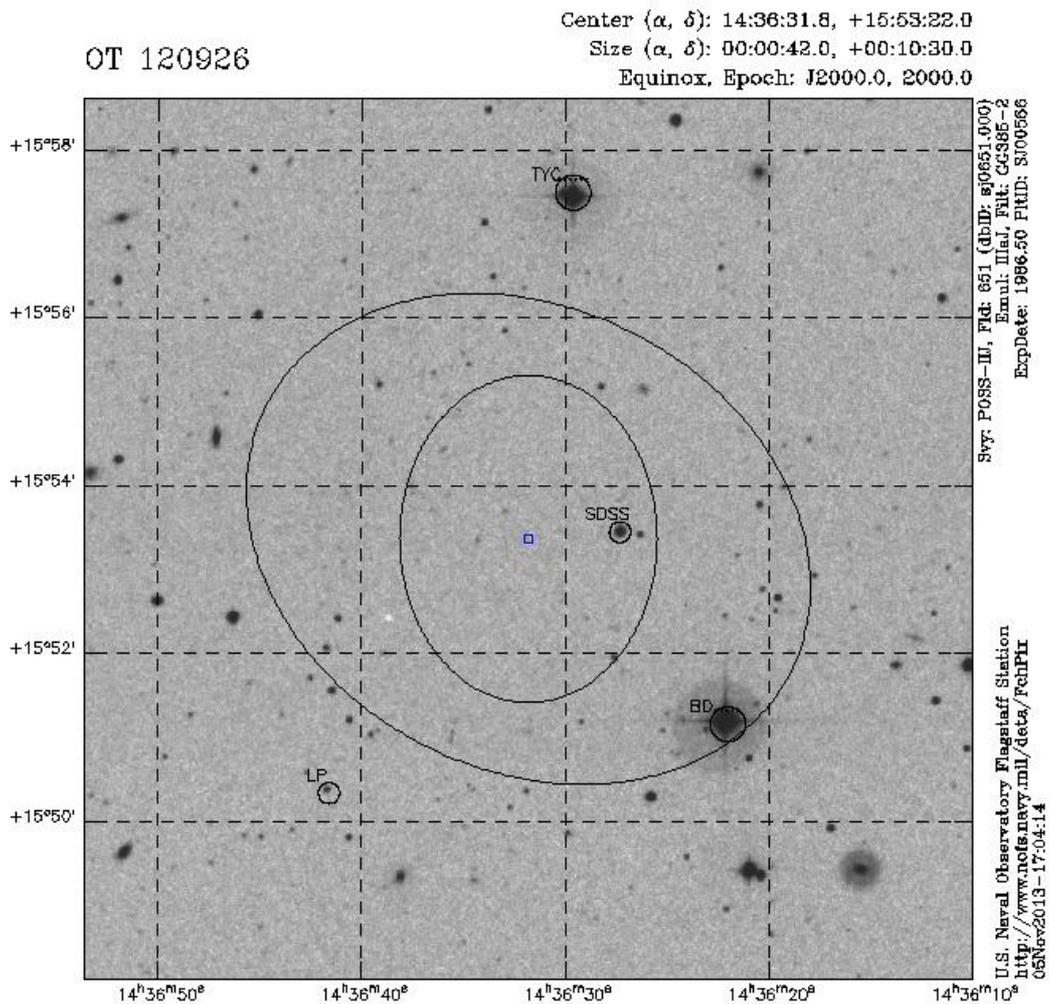

Figure 5: The box on this finding chart shows the weighted average position of OT 120926. The inner ellipse shows the 3-sigma random uncertainty on that position. The outer ellipse incorporates both the 3-sigma random uncertainty and the 3-sigma systematic uncertainty (from the uncertainty on the pixel scale) on the position of OT 120926, added together linearly (not in quadrature). The four objects discussed in the text as possible counterparts are labeled by their IAU prefix.

We used the databases of the CRTS (Catalina Real-time Transient Survey; Drake et al. 2009) and the ASAS (All Sky Automated Survey; Pojmanski 1997) to search for photometric light curves of these objects.

The object SDSS J1436+1553 (SDSS J143627.19+155326.8 = ASAS 143627+1553.5) was taken into consideration because of its relative proximity to OT 120926. In the CRTS, this object has a constant magnitude around 14.0, which is consistent within the errors with the result given by the ASAS.

The red, high proper motion star LP 440-48 (Luyten 1981) has a magnitude of 15.9 in the CRTS, with no sign of variability. It is not found in the ASAS.

When comparing the photometry results from the CRTS and ASAS on the objects BD +16 2617 and TYC 1477-341-1, there emerged a large discrepancy. There are no conspicuous changes in their magnitudes judging from the ASAS light curves, but the CRTS results suggest three to four magnitudes of variability in both objects. In fact, CRTS photometry data for these objects are untrustworthy because the objects are often saturated.

We also searched for variable objects within a radius of 5 arcmin around OT 120926. We found one object (CSS J143625.1+155102) with apparent variability which, however, is due to its proximity to a much brighter star. There are no useful data available in ASAS for this object.

We examined cutouts of the CRTS images at the location of OT 120926 (kindly provided by A. Drake), but found no evidence for an uncatalogued variable object in them.

Examination of the POSS-I and POSS-II plates within a 6 arc minute radius of the transient's coordinates did not reveal any objects with dramatic variability. (Note that the POSS-I O plate has at least six point-like defects within a 6 arc minute radius which are spurious, as they do not correspond to the positions of any objects in the SDSS images of the region.)

Examination of GALEX images of the field did not reveal any objects with unusual ultraviolet-optical colors. The star BD +16 2617 is much brighter in the UV than TYC 1477-341-1, but both stars have UV-optical colors consistent with expectations for stars with their optical colors. The star LP 440-48 is not detected by GALEX, and the star SDSS J1436+1553 is barely detected.

Examination of the HEASARC X-ray and gamma-ray satellite databases did not reveal any X-ray sources near the position of the transient, nor any gamma-ray burst consistent with its location and time of appearance.

Conclusion:

The optical transient OT 120926 flared to naked-eye brightness for at least half an hour, but was not at naked-eye brightness several hours before the observation. Database searches at its position yield no unambiguous identification of a quiescent counterpart of this transient, but do identify several candidates. A flare on the high proper motion, probable M dwarf star LP 440-48 could have produced OT 120926, but the amplitude of the flare would be an unprecedented 11.3 magnitudes, as compared to the previous record amplitude of 9.5 magnitudes (Schaefer 1990). OT 120926 could be an outburst from a previously unrecognized cataclysmic variable identified with LP 440-48 or one of two brighter stars in our error circle. If the latter is the case, the outburst would be 6.8 to 7.3 magnitudes, consistent with the known range of CV outburst amplitudes (Harrison et al. 2004). However, none of the candidate stellar counterparts of OT 120926 have shown any credible evidence of previous variability in the ASAS or CRTS. Spectroscopy of the possible counterparts of OT 120926 is needed as a next step to identifying its quiescent counterpart and determining the nature of this remarkable event.

Acknowledgements:

We thank K. Stanek, A. Drake, C. Kochanek, and J. L. Prieto for archival searches and suggestions, P. Lloyd for initial conversations regarding this object, Joe Kristl for all-sky images from West Acton Observatory, and the referee for a thorough and speedy review.

Y. Zhao was supported at York University by the MITACS Globalink program. P. Hall thanks NSERC for research support.

The Liverpool Telescope is operated on the island of La Palma by Liverpool John Moores University in the Spanish Observatorio del Roque de los Muchachos of the Instituto de Astrofisica de Canarias with financial support from the UK Science and Technology Facilities Council. Kitt Peak National Observatory, National Optical Astronomy Observatory, is operated by the Association of Universities for Research in Astronomy (AURA) under cooperative agreement with the National Science Foundation. The Image Reduction and Analysis Facility (IRAF) is also distributed by NOAO. MMT Observatory is a joint facility of the University of Arizona and the Smithsonian Institution.

Tables:

Table 1

| Image Name | UTC of Observation | Exposure Time | Pixel Scale (arcsec/pixel) |
|---|---|---|---|
| DSC1861 | 2012-09-26 00:17:54 | 2 sec | 146.49±1.94 |
| DSC1875 | 2012-09-26 00:45:50 | 0.125 sec | 83.67±1.74 |

Table 2

|  | RA/degrees | DEC/degrees | RA | DEC |
|---|---|---|---|---|
| DSC1861 | 219.1299±0.0122 | 15.8928±0.0122 | 14:36:31.2±0:02.9 | +15:53:34±0:44 |
| DSC1875 | 219.1359±0.0116 | 15.8778±0.0232 | 14:36:32.6±0:02.8 | +15:52:40±1:24 |
| Average | 219.1331±0.0084 | 15.8896±0.0108 | 14:36:31.9±0:02.0 | +15:53:22±0:39 |

Table 3

| Image | UTC | Magnitude |
|---|---|---|
| Liverpool | 2012-09-25 19:51:14-20:00:03 | >5 |
| DSC1861 | 2012-09-26 00:17:54 | 4.7±0.2 |
| DSC1875 | 2012-09-26 00:45:50 | 5.6±0.3 |

Table 4

| Object | V Magnitude | Distance (arcsec) | Distance (sigma) | RA | DEC |
|---|---|---|---|---|---|
| OT 120926 | 4.70 | 0.00 | 0.00 | 14:36:31.9±0:02.0 | +15:53:22±0:39 |
| SDSS J1436+1553 | 13.99 | 70.81 | 2.05 | 14:36:27.19 | +15:53:26.83 |
| BD +16 2671 | 11.49 | 132.90 | 3.85 | 14:36:21.97 | +15:51:09.47 |
| LP 440-48 | 15.96 | 182.06 | 5.27 | 14:36:41.59 | +15:50:20.20 |
| TYC 1477-341-1 | 12.03 | 247.51 | 7.16 | 14:36:29.56 | +15:57:29.50 |